\documentclass[osajnl,preprint,showpacs,superscriptaddress,12pt]{revtex4-1} 
\usepackage{amsmath,amssymb,graphicx}
\usepackage{textcomp}
\usepackage{color}
\begin{document}

\title{Single shot three-dimensional imaging of dilute atomic clouds}

\author{Kaspar Sakmann}\email{Corresponding author: sakmann@stanford.edu}
\author{Mark Kasevich}
\affiliation{Department of Physics, Stanford, California, USA}

\begin{abstract}
Light field microscopy methods together with three dimensional (3D) deconvolution can be used to obtain 
single shot 3D images of atomic clouds.   We demonstrate the method using a test setup which extracts three 
dimensional images from a fluorescent $^{87}Rb$ atomic vapor. 
\end{abstract}


\maketitle 

In the field of ultracold atoms  dilute atomic clouds 
are usually imaged at the end of an experimental run 
either by fluorescence or absorption imaging. 
In both cases, the observed images consist of projections of a three dimensional 
(3D) density distribution on an imaging plane.  In many such experiments, 
a full three dimensional reconstruction would be useful.  
For example, in the atom interferometry work of Ref. \cite{Alex13}, 
a 3D image at the output of the interferometer would enable direct 
volumetric extraction interferometer phase shifts 
(thus providing information about the rotation and acceleration of the apparatus). 


Here, we show that light field imaging 
developed in the computer vision community \cite{Adelson, Ng,Levoy} 
can be successfully used to obtain 3D images of fluorescent clouds of
atoms in single shot measurements. Moreover, the light field imaging technique we use is highly adaptable to
typical imaging systems in such experiments. The main modification
consists of placing a microlens array  in the optical path.
By recording the light field emitted from the atomic cloud, 
a stack of focal planes can be obtained by refocusing computationally. 
Furthermore, by recording the point spread function (PSF) of the imaging system, this focal stack
can be used as a starting point for 3D deconvolution as shown in \cite{Levoy}. The light field imaging technique used here leads to a 
reduced transverse resolution compared to a conventional imaging setup 
\cite{Levoy}.
For many ultracold atom experiments this loss of resolution is acceptable though.
For example, the atomic clouds in atom interferometers can measure almost a centimeter and 
show only a small number of equally distanced fringes over this length scale \cite{Alex13,Susannah13}.

In the following we briefly review the concepts of light 
field microscopy as far as they are relevant to this work. A complete treatment can be found in \cite{Levoy}.
We then demonstrate light field microscopy of fluorescent $^{87}$Rb with 3D resolution in a test setup.
We measure the PSF of this imaging system and apply a 3D deconvolution algorithm to the stack of focal planes. 
The deconvolution improves the quality of the focal stack substantially. 

\begin{figure}[htbp]
\centerline{\includegraphics[width=0.7\columnwidth]{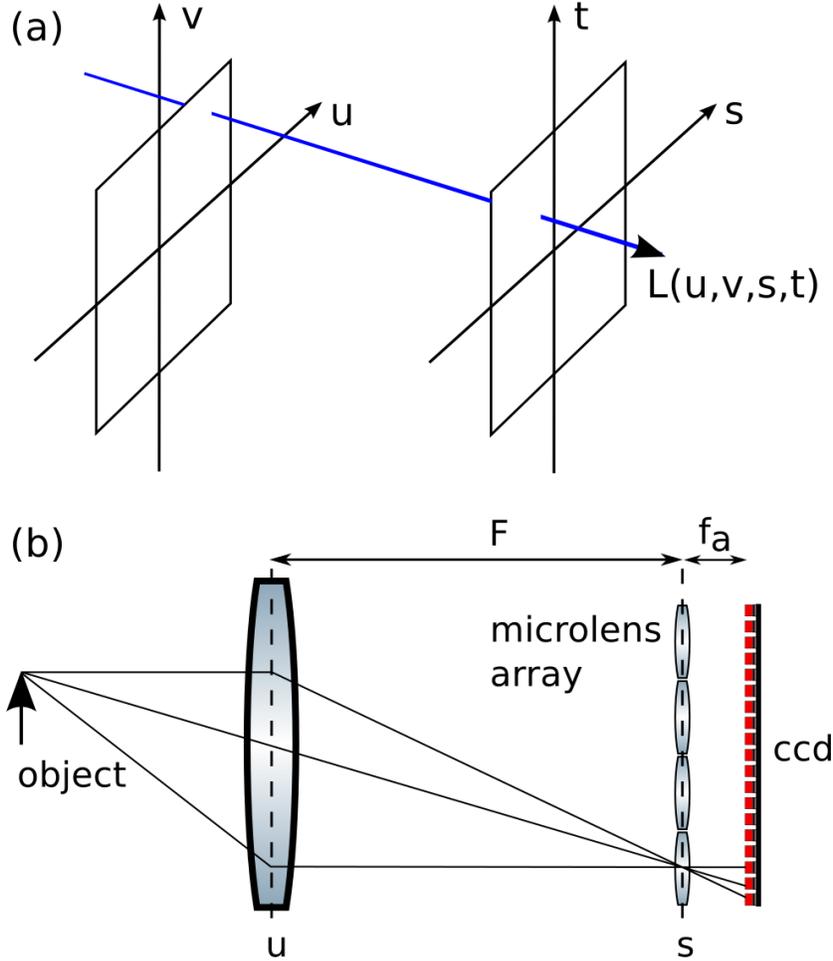}}
\caption{Principle of light field imaging. 
(a) A ray of light intersects two planes separated by a distance $F$ at locations $(u,v)$ and $(s,t)$ 
Its radiance is denoted by $L_F(u,v,s,t)$. The collection of all such rays is called the light field.
(b) An object is imaged by a main lens onto a microlens array. The CCD chip behind the array records the light field.
}\label{F1}
\end{figure}
The principle of a light field is illustrated in Fig. \ref{F1} (a). 
A ray of light emitted from an object can be parameterized by its intersections with two parallel planes, separated by a distance $F$ from each other. 
The radiance of a ray intersecting the two planes at locations $(u,v)$ and $(s,t)$ is denoted by $L_F(u,v,s,t)$. 
All such rays together are called the light field of the object. 
The light field is recorded by the setup shown in Fig. \ref{F1}(b):
a main lens forms an image of an object at a distance $F$, where an array of 
microlenses is located. A CCD camera is located one microlens focal length $f_a$ behind the array. 
The location of the main lens and the microlens array define the $(u,v)$ and the $(s,t)$ plane, respectively. 
The different pixels on the CCD behind the microlens located at $(s,t)$ record 
the radiance coming from all different locations $(u,v)$ on the main lens. 
Thus, the CCD camera samples the light field $L_F(u,v,s,t)$. 
The microlenses act like pinholes in this setup and the $f$-number $N_a$ of the microlens array 
must be smaller than the image side $f$-number $N_{obj}$ of the main lens \cite{Ng}. For optimal usage of the CCD chip 
the $f$-numbers should be matched.  
In a conventional camera a CCD chip is located in the plane of the microlens array recording the irradiance
\begin{equation}\label{irradianceF}
E_F(s,t)=\frac{1}{F^2} \int\int L_F(u,v,s,t)du dv.
\end{equation}
For simplicity we neglect an illumination falloff factor $\cos(\theta)^4$ in (\ref{irradianceF}) which describes vignetting of rays that form large angles $\theta$
with the CCD array (in our setup $\cos(\theta)^4\ge 0.995$ and this approximation is justified).
The resolution --assuming ray optics-- is then determined by the size of the pixels on the CCD array. 
For light field imaging the double integral (\ref{irradianceF}) is evaluated by summing up the
values of all CCD pixels under the microlens centered at $(s,t)$, see Fig. \ref{irradianceF}(b). 
The lateral resolution is then determined by size of a microlens, i.e.reduced when compared to a conventional camera. 
However, the recorded light field allows the computation of the irradiance 
$E_{\alpha F}$ at planes located at distances $\alpha F$ behind the main lens with $\alpha\neq 1$ \cite{Ng}:
\begin{eqnarray}\label{irradianceaF}
E_{\alpha F}(s,t)=\frac{1}{\alpha^2F^2} \times \quad\quad \quad\quad \quad\quad \quad\quad \quad\quad \quad\quad \quad\quad \quad\quad  \nonumber\\ \int\int L_F\left(u,v,u(1-\frac{1}{\alpha})+\frac{s}{\alpha},v(1-\frac{1}{\alpha})+\frac{t}{\alpha}\right)du dv.\qquad
\end{eqnarray}
Equation (\ref{irradianceaF}) forms the basis for computational refocusing. 
\begin{figure}[htbp]
\centerline{\includegraphics[width=0.7\columnwidth]{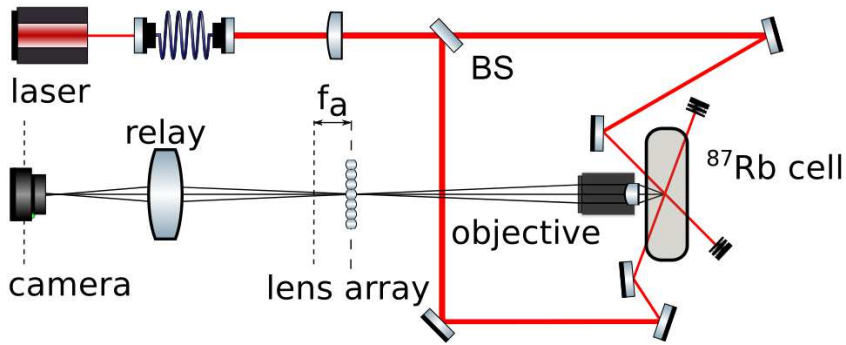}}
\caption{Schematic of the experiment. A laser beam  is focused by a lens and 
split by a beam splitter. The beams intersect in a $^{87}Rb$ vapor cell at different angles relative to the imaging direction. 
The fluorescence is imaged using a microscope objective, a microlens array and a relay lens
onto a CCD camera.}\label{F2}
\end{figure}

In the following we briefly discuss our experimental test setup for 3D imaging of dilute atomic clouds. 
Fig. (\ref{F2}) shows a schematic. A Gaussian laser beam resonant with the $5 {^2}S_{1/2} \rightarrow 5 {^2}P_{3/2}$ 
transition of $^{87}$Rb ($\lambda=780nm$) passes a focusing lens and is split by a beam splitter. 
The two beams intersect at their waists in a $^{87}$Rb vapor cell.  
The  $1/e^2$ waist of each beam is about $55\mu m$ ($w_0=27.5\mu m\pm1\mu m$) corresponding to a 
Rayleigh range $b=2z_0=6mm\pm0.2mm$. The fluorescence of the $^{87}$Rb atoms is imaged 
using a $10X/NA=0.25$ Semi-Plan Objective (Edmund Optics). 
A magnified image, $M_1=-11$, of the intersecting beams is formed in the plane of the microlens array.  
Thus, the image side $f$-number of the objective is given by $N_{obj}=\vert M_1\vert/(2NA)=22$.
The microlens array (RPC Photonics) has a pitch of $\Delta=125\mu m$ and a focal length of $2.5mm$, corresponding to an $f$-number of $N_a=20$. 
The slight mismatch of the $f$-numbers ensures that image circles coming from adjacent microlenses remain separated on the CCD chip. 
The focal plane of the microlens array is relayed at a magnification of $M_2=-0.54$ onto a $1/3''$ CCD chip. 
The camera uses $1280 \times 960$ pixels of the CCD chip, each $3.75\mu m$ in size.  
The number of pixels behind each microlens is very close to $18\times18$.
After cropping the image to an integer number of lenslets ($71\times53$)
the field of view of this setup is  $800 \mu m \times 600 \mu m$ in object space. 
Each lenslet of the array therefore covers an area of $11.3\mu m\times11.3\mu m$. 
\begin{figure}[htbp]
\centerline{\includegraphics[width=0.7\columnwidth]{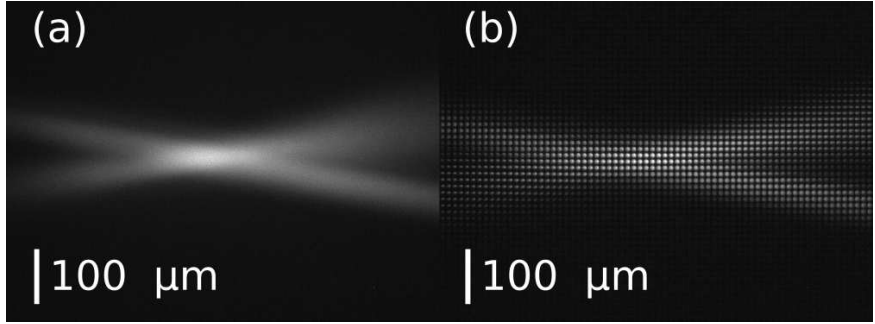}}
\caption{ Fluorescence of $^{87}$Rb atoms in a vapor cell. (a) 
Conventional image of two focused laser beams intersecting in a $^{87}Rb$ vapor cell at their waists. 
(b) Light field recording of the same scene using a microlens array. The light field is extracted from the
individual pixels behind each microlens.}\label{F3}
\end{figure}

Fig. \ref{F3}(a) shows a conventional image of the fluorescence of $^{87}$ Rb atoms 
illuminated by two laser beams intersecting in the vapor cell. The resolution is higher than needed: the object has no 
features smaller than the beam waist of $55\mu m$. 
As shown in Fig. \ref{F2} the laser beams do not lie in a plane perpendicular to the imaging axis, which we call the $z$-direction from now on.
However, it is not possible to determine the $z$-components of the laser beams from Fig. \ref{F3}(a). 
Fig. \ref{F3}(b) shows a light field recording of the same scene using a microlens array, as discussed above. $71\times53$ microlenses of the array
discretize the $(s,t)$ plane. The $(u,v)$ plane is discretized by $18\times18$ pixels behind each microlens.
In total this provides a light field $L_F(u,v,s,t)$ sampled at $71\times53\times18\times 18$ points. We note that evaluating the
double integral (\ref{irradianceaF}) requires the interpolation of $L_F$ in the variables $s$ and $t$.

\begin{figure}[htbp]
\centerline{\includegraphics[width=0.7\columnwidth]{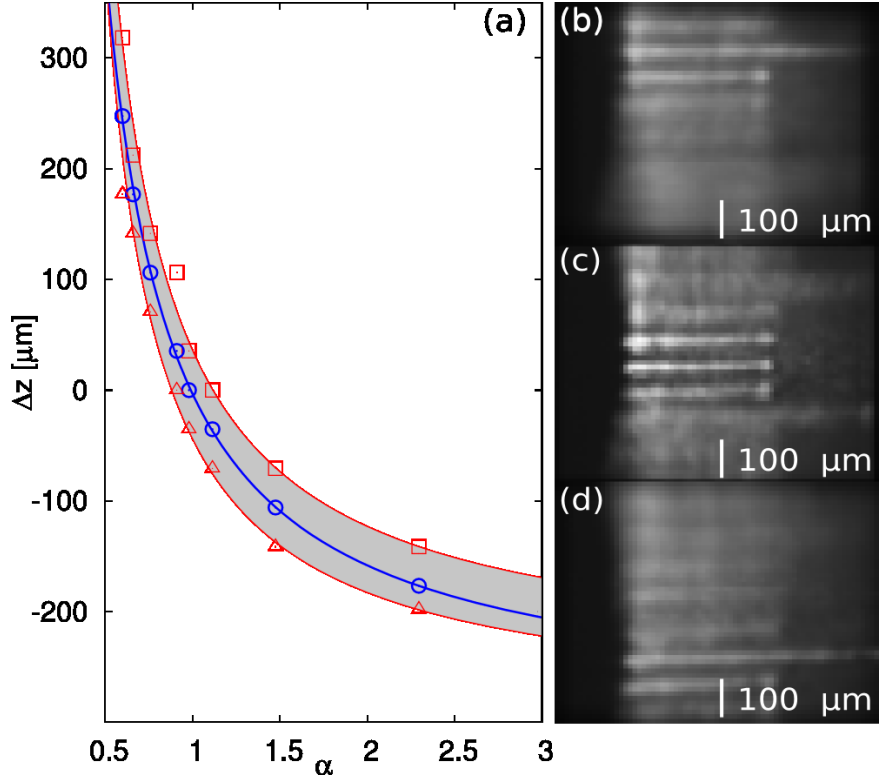}}
\caption{Calibration of the imaging setup. The test target is a ruler (mark spacing $100\mu m$)
tilted at an angle of 45\textdegree with respect to the imaging axis. 
Its light field is recorded and the image is computationally refocused to different object planes
located at $\Delta z$. The values of $\alpha$ corresponding to the sharpest contrast are recorded 
together with the depth of field.   
(a) shows measurements of $\Delta z$ as a function of $\alpha$ ($\bf \color{blue}{\odot}$) as well as the corresponding 
depth of field  (${\bf \color{red}{\Box}}$ and $\color{red}{\bigtriangleup}$). Also shown are fits of the measurements 
to the thin lens equation (solid lines). The shaded  area represents the depth of field. (b-d) 
show refocused images of the ruler. From top to bottom $\alpha=0.60,0.86$ and $1.77$. 
}\label{F4}
\end{figure}

In principle, the obtained light field allows the evaluation of (\ref{irradianceaF}) for a given 
$\alpha$ and thereby refocusing to different image planes.  
A calibration measurement is needed though to determine the location of the 
object planes corresponding to a given $\alpha$.
3D deconvolution in particular requires equally spaced planes in object space and also knowledge of the PSF.
As refocussing is performed computationally, 
it is necessary to image a test target with known depth information for calibration.
Here, we use a ruler tilted at an angle of 45\textdegree with respect to the $z$-axis.
The marks on the ruler are spaced by $100\mu m$. We record the light field of this ruler, 
refocus computationally to every mark on it and note the values of $\alpha$ corresponding to the 
sharpest contrast as well as the depth of field. Since the 
paraxial approximation is still well satisfied (at NA=0.25 the error is only $\approx2\%$)
the entire imaging system can be described by an 
effective thin lens equation $1/f=1/s_o+1/s_i$ with a focal length $f$, an object distance  $s_o=z_0+\Delta z$ and an 
image distance $s_i=\alpha F$. $z_0$ denotes the object distance corresponding to $\alpha=1$.
Measurements of the sharpest contrast and the depth of field are shown together with fits to the thin lens equation in Fig. \ref{F4} (a).
This calibrates the imaging setup.
Examples of refocused images of the ruler are shown in Fig. \ref{F4} (b-d). 

We now go on to determine the 3D PSF by imaging a pinhole of $d=1\mu m$ diameter. The numerical aperture of the pinhole's Airy disk is determined by 
$\sin\theta=1.22\lambda/d\approx0.95$ which is much greater than the numerical aperture of the microscope objective. 
At the given magnification of $M_1=11$ and microlens array pitch of $125\mu m$ its image is contained in a single microlens and the pinhole 
approximates a subresolution isotropic point source, as is required for determining the PSF.
We note that microscope objectives are object-side telecentric and therefore produce orthographic views. As a consequence 
the PSF becomes independent of position in the plane orthogonal to the $z$-direction, the $xy$ plane. 
This allows the use of a single PSF, shift invariant in the $xy$ plane. For more details, see \cite{Levoy}.  
The light field of the pinhole determines the PSF of the imaging setup: refocusing to 
different object planes provides the 3D structure of the PSF. 
The intensity profile is very well approximated by a cylinder symmetric 2D Gaussian.
In order to avoid asymmetries in the PSF stemming from image noise we use the Gaussian fit 
to the intensity profile for refocusing.
The resulting 3D PSF is well described by the following model PSF 
\begin{figure}[htbp]
\centerline{\includegraphics[width=0.6\columnwidth]{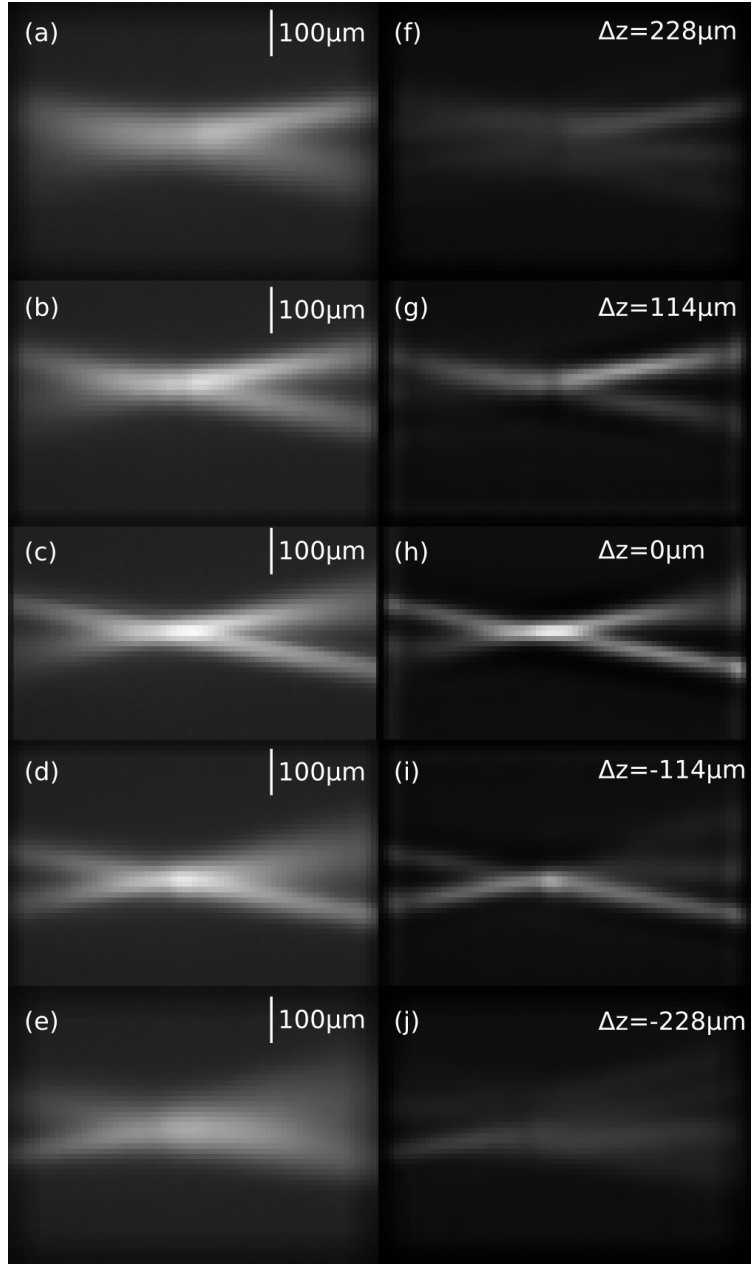}}
\caption{Refocusing of the fluorescent $^{87}$Rb atoms shown in Fig. \ref{F3}. Subfigures  (a-e) show slices of a focal stack obtained by computational refocusing 
the light field shown in Fig. \ref{F3}(b). Contrary to Fig. \ref{F3}(a) it can now be seen that 
the laser beams intersect at an oblique angle with respect to the imaging axis. Nevertheless, the images show a considerable amount of blur.
Subfigures (f-j) show the result of the deconvolution of the focal stack with the measured point spread function using the Richardson-Lucy algorithm. 
The blur is strongly reduced and the 3D structure becomes clearly visible.}\label{F5}
\end{figure}
\begin{equation}\label{PSF}
\Psi(x,y,z)=\frac{1}{\sigma(z)^2} e^{-\frac{x^2+y^2}{2\sigma(z)^2}}
\end{equation}
with $\sigma(z)$ increasing linearly with $\vert z\vert$ starting from a minimal value $\sigma_{min}$ at $z=0$.  
The PSF model (\ref{PSF}) ensures that the luminescence is conserved for every $z$ plane.

We now turn to refocusing the fluorescence of the $^{87}$Rb atoms shown in Fig. \ref{F3}(a). 
First we use the light field shown in Fig. \ref{F3}(b) as well as 
the calibration data to obtain a focal stack of the image of the two laser beams
spaced by $\sim13\mu m$ in the $z$-direction. Some of the slices are shown in Fig. \ref{F5}(a-e). 
In contrast to the conventional image shown in Fig. \ref{F3}(a) the 3D structure of the scene 
can now be deduced from the focal stack: the laser beams can be seen to be directed at an angle with respect to 
the $xy$ plane and the parts of the laser beams shown in the bottom half of Fig. \ref{F3}(a) are obviously closer to the objective 
than those in the top half. 

While refocusing alone provides the overall 3D structure, each of the images in Fig. \ref{F5}(a-e) is 
considerably blurred. For example the region where the beams intersect is about $100\mu$ wide much wider than the beam waist of $55\mu m$.
However, the obtained focal stack together with the PSF allow for 3D deconvolution techniques to be applied. The recorded image of an object is 
the convolution of the object with the PSF of the imaging setup. Techniques that invert this operation are called 
deconvolution algorithms \cite{Blahut}. Here, we use the Richardson-Lucy algorithm which takes the PSF of the imaging system and a focal stack
in order to obtain a maximum likelihood estimate of the object \cite{Blahut}. The result of deconvoluting the focal stack 
using the Richardson-Lucy algorithm is shown in Fig. \ref{F5}(f-j). The results are compelling:  as $\Delta z$ is 
scanned from about $-230\mu m$ to about $230\mu m$
it can clearly be seen how the two laser beams enter the field of view from below, 
intersect near $\Delta z=0$, separate and finally leave the field of view again. Moreover, it can also be 
seen that the laser beam coming from the left makes a steeper angle with the 
$xy$ plane than the laser beam coming from the right, in agreement with the experimental setup, as shown in Fig. \ref{F2}.
The deconvoluted focal stack also agrees quantitatively with the expected beam waist size: the spatial extent of the intersection of the beams 
is now about the same as the expected $55\mu m$ of each of the beams.

Summarizing, we have demonstrated that light field microscopy techniques combined with 3D deconvolution can be successfully used to 
obtain the 3D structure of clouds of fluorescent $^{87}Rb$ atoms, which are commonly used in ultracold gases experiments. 
We expect these methods to be enabling for cold and ultra-cold atom experiments 
where fluorescence detection is employed and three dimensional information on 
the spatial distribution of the atom cloud is desired.

We acknowledge help by S.-W. Chiow and J. Hogan during the initial phase of the experiment, 
as well as discussions with M. Levoy and M. Broxton.
K. S. acknowledges funding through the Karel Urbanek Postdoctoral Research Fellowship.

%




\end{document}